\documentclass{appolb}
\usepackage{epsfig}
\usepackage[utf8]{inputenc}
\usepackage{graphicx}
\usepackage{caption}
\usepackage{dcolumn}
\usepackage{bm}
\usepackage{color}

\newcommand{\pip}{$\pi^{+}$}
\newcommand{\pim}{$\pi^{-}$}
\newcommand{\kap}{$K^{+}$}
\newcommand{\kam}{$K^{-}$}

\newcommand{\sNN}{$\sqrt{s_{\rm NN}}$}
\newcommand{\be}{\begin{equation}}
	\newcommand{\ee}{\end{equation}}

\def\muB{$\mu_B$}

\begin{document}
	\title{Collisions of Small Nuclei in the Thermal Model}
\thanks{Presented at Critical Point and Onset of Deconfinement CPOD2016, University of Wroc\l aw, Poland , May 30th - June 4th 2016}%
	\author{J.~Cleymans%
	\address{UCT-CERN Research Centre and Department  of  Physics,\\ University of Cape Town, Rondebosch 7701, South Africa}
\and
	B.~Hippolyte
	\address{Institut Pluridisciplinaire Hubert Curien (IPHC), Universit\'e de Strasbourg, CNRS-IN2P3, Strasbourg, France}
\and	
H.~Oeschler
        \address{Physikalisches Institut, Ruprecht-Karls-Universit\"at Heidelberg, Heidelberg, Germany}
\and
K.~Redlich
	\address{Institute of Theoretical Physics, University of Wroc\l aw, Pl-45204 Wroc\l aw, Poland}
	\address{ExtreMe Matter Institute EMMI, GSI, D-64291 Darmstadt, Germany}
\address{Department of Physics, Duke University, Durham, NC 27708, USA}
\and
	N.~Sharma
	\address{Department of Physics, Panjab University, Chandigarh 160014, India}
}
\maketitle
\newpage
\begin{abstract}
An analysis is presented of the expectations of the thermal model
for particle production in collisions of small nuclei. 
The  maxima observed in particle ratios of strange particles to pions as 
a function of beam energy in heavy ion collisions, are reduced when considering smaller nuclei.
Of particular interest is
the $\Lambda/\pi^+$ ratio shows the strongest  maximum which survives even in collisions of small nuclei.

\end{abstract}


	\date{\today}

\section{Introduction}
A large effort is presently under way to study not only heavy- but also light-ion
collisions. This is being motivated by the results obtained in heavy ion collisions like Pb-Pb and Au-Au,  for
 the \kap/\pip,  and also other particle ratios. It  
has been  conjectured that these indicate a phase change  in nuclear matter~\cite{Gazdzicki:1998vd}.

A consistent description of particle production in heavy-ion collisions, up to LHC energies,
has emerged during the past two decades using a thermal-statistical model (referred to simply as thermal model in the remainder of this talk). 
It is based on the creation and subsequent decay of hadronic resonances produced in chemical
equilibrium at  unique temperature and baryon chemical potential.  According to this picture the bulk
of hadronic resonances made up of the light flavors u,d and s valence quarks are  produced in chemical equilibrium. 

Indeed some particle ratios exhibit very interesting features when studied as a function of the beam energy  which deserve 
attention: (i) a maximum in the \kap/\pip ratio, (ii) a maximum in the $\Lambda/\pi$ ratio, (iii) no maximum in the \kam/\pim ratio.
The maxima occur  at a center-of-mass energy of around 10  GeV~\cite{Andronic:2005yp,Cleymans:2005xv,Cleymans:2004hj}. 
It is interesting to note that the occurrence of these maxima happens in an energy regime where a maximum 
baryon density occurs~\cite{Randrup:2006nr} and a transition from baryon-dominated freeze out to a 
meson dominated one takes place~\cite{Cleymans:2004hj}. An alternative interpretation
is that these maxima reflect a phase change~\cite{Gazdzicki:1998vd} to deconfined state of matter.

The maxima mark a distinction between heavy-ion collisions and p-p collisions  as  they are not observed in the latter. 
This shows a clear difference between the two systems which is worthy of further investigation.

It is the purpose of the present talk to report on an analysis~\cite{boris} studying  the transition from a small system like a p-p collision to a large system
like a Pb-Pb or Au-Au collision and to follow  explicitly the genesis of the maxima in certain particle ratios. 
\section{The model}
A relativistic heavy-ion collision will go through several stages. 
At one of the later stages, the system will be dominated by
hadronic resonances. 
The identifying feature of the thermal model  is that all the resonances as listed by the Particle Data Group~\cite{Agashe:2014kda} are 
assumed to be in thermal and chemical equilibrium.
This  assumption drastically reduces the number of free parameters and thus this stage is determined by just a few 
thermodynamic variables namely, the chemical freeze-out temperature $T$, the various chemical potentials $\mu$ determined by 
the conserved quantum numbers and by the volume $V$ of the system. The latter plays no role when considering ratios
of yields. It has been shown that this description is also the correct 
one~\cite{Cleymans:1997eq,redlich,Akkelin:2001wv,Broniowski:2001we} for a scaling expansion as first discussed by Bjorken~\cite{Bjorken:1982qr}.

In general, if the number of particles carrying quantum numbers related to a conservation law is small, then
the grand-canonical description no longer holds. In such a case conservation of quantum numbers has to be implemented
exactly in the canonical ensemble~\cite{BraunMunzinger:2001as,BraunMunzinger:2003zd}. 
In the case considered here  there are two volume parameters: the overall
volume of the system $V$, which determines the particle yields at fixed density and the
strangeness correlation (cluster) volume $V_c$, which reflects the canonical
suppression factor and reduces the densities of strange particles. 
Assuming spherical geometry, the volume $V_c$ is parameterized by the radius $R_C$ which serves as a free
parameter and defines the range of local strangeness equilibrium. 
\section{Origin of the maxima}

According to the thermal model, the baryon chemical potential decreases continuously with increasing beam energy.
At the same time the temperature increases rather quickly until it reaches a plateau.
Following the rapid rise of the temperature at low beam energies, the $\Lambda/\pi^+$ 
and  \kap/\pip also increase rapidly.
This halts when the temperature reaches its limiting value. However, simultaneously the baryon chemical potential keeps on decreasing.
Consequently, the  $\Lambda/\pi$ and \kap/\pip ratios follow this decrease due to
 strangeness conservation as \kap~is produced in associated production together with a $\Lambda$.  
The two effects combined lead to maxima in both cases. 
For very high energies, the baryo-chemical potential no longer plays a role ($\mu_B\approx 0 $) and the temperature is constant
hence these ratios hardly vary~\cite{Cleymans:2004hj}.

\begin{figure}[h]
\includegraphics[width = 0.46\textwidth]{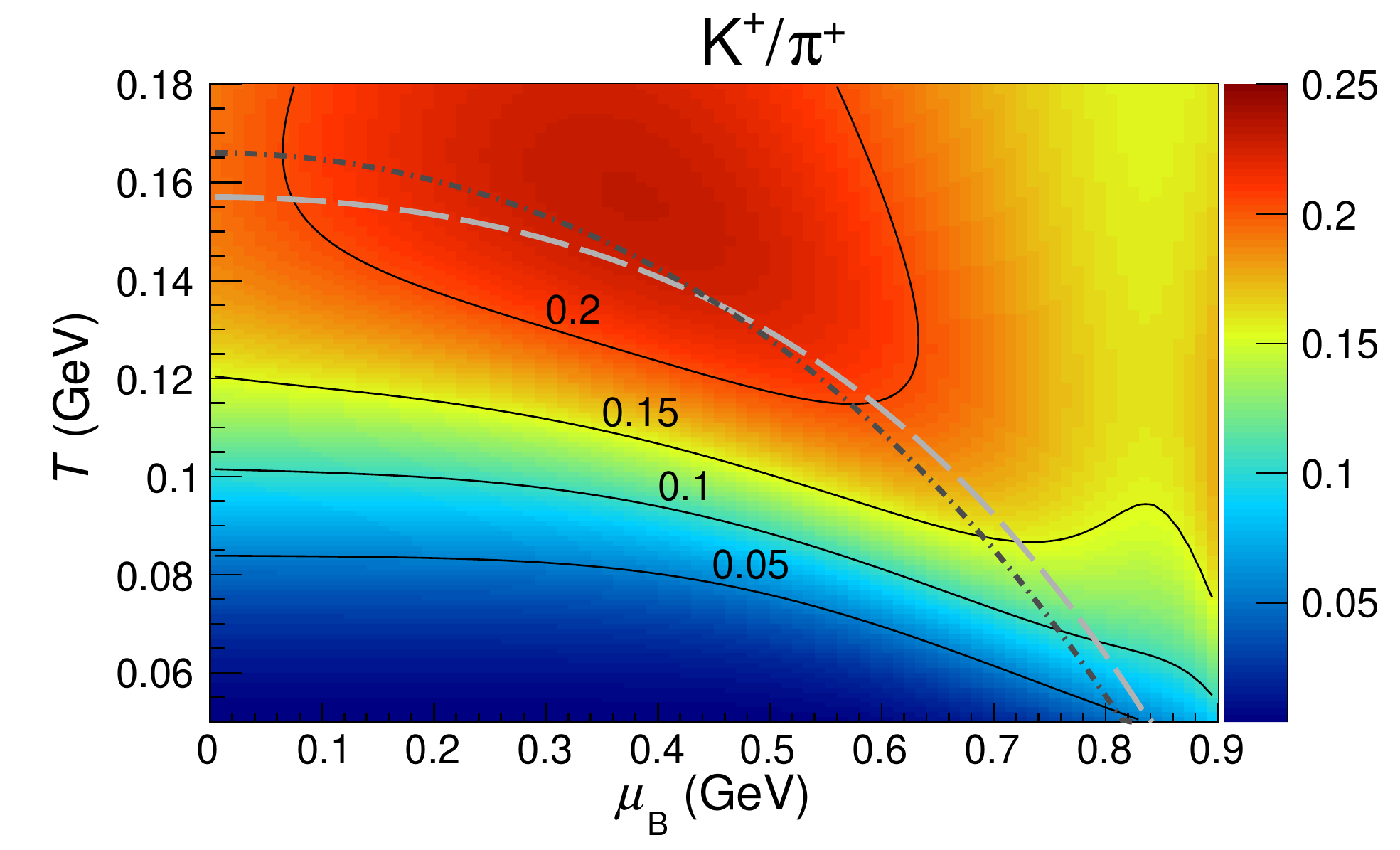} 
\hfill\includegraphics[width = 0.46\textwidth]{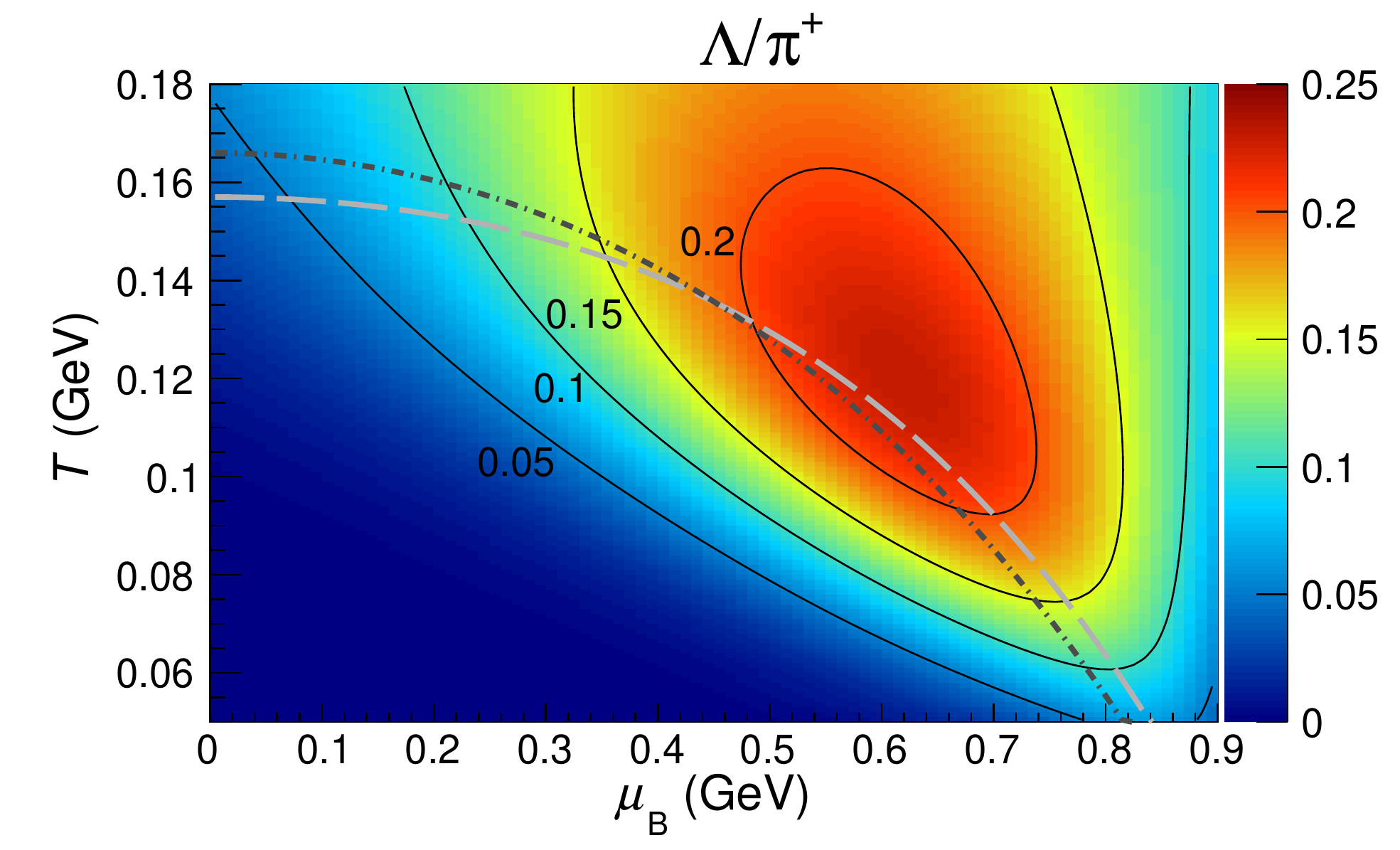} 
\caption{Values of the \kap/\pip (left-hand pane) and the $\Lambda/\pi^+$ (right-hand pane) ratios in the $T-\mu_B$ plane.
Lines of constant values are indicated. 
The dashed-dotted line is the freeze-out curve obtained in~\cite{Cleymans:2005xv} while the dashed line 
uses the parameterization given in~\cite{Vovchenko:2015idt}.
Note that the maxima do not occur in the same position.
 }
\label{fig:constant}
\end{figure}

To show this in more detail we present as an example in Fig.~\ref{fig:constant}   lines where the \kap/\pip  and the 
$\Lambda/\pi^+$ ratios remain constant in the $T-\mu_B$ plane.  
It should be noted that the  maxima of these ratios do not  occur in the
same position, which remains to be confirmed experimentally. It is also worth noting that the maxima 
are not on but slightly above the freeze-out curve. 

\section{Particle ratios for small systems}

To consider the case of the collisions of smaller nuclei we have to take into account
the strangeness suppression according to the canonical model, i.e.~~the concept of strangeness correlation 
in clusters of a sub-volume $V_c\leq V$~\cite{Cleymans:1998yb,Hamieh:2000tk,Kraus:2007hf}. 

A particle with
strangeness quantum number $s$ can appear anywhere in the volume $V$ but it has to
be accompanied by another particles carrying strangeness $-s$
to conserve strangeness in the correlation volume $V_c$ .
Assuming spherical geometry, the volume $V_c$ is parameterized by the radius $R_C$ which is a free
parameter that defines the range of local strangeness equilibrium. 

In the following we show the trends of various particle ratios as a function of \sNN.  
The  dependence of $T$ and \muB{}  on the beam energy is taken from  heavy-ion collisions~\cite{Cleymans:2005xv}. 
For p-p collisions slightly different parameters are more suited~\cite{Cleymans:2011pe}. 
Therefore, the calculations shown give the general trend. 
We have ignored the variations of other parameters with system size. 

We focus on the system-size dependence of the thermal parameters with particular emphasis on
the change in the strangeness correlation radius $R_C$. The parameters $R$ = 10 fm (which is
the value for central Pb-Pb collisions) and $\gamma_S$ = 1  are kept fixed. The freeze-out values
of $T$ and $\mu_B$ will vary with the system size~\cite{Kraus:2007hf}, however this has not been taken 
into account in the present work
which aims to give a qualitative description of the effect.

The smaller system size is described by decreasing the value of the correlation radius $R_C$. 
This ensures that strangeness
conservation is exact in $R_c$, and that strangeness production is suppressed with
decreasing $R_c$. 

\begin{figure}[h]
\hspace{-0.2cm}\includegraphics[width = 0.5\textwidth]{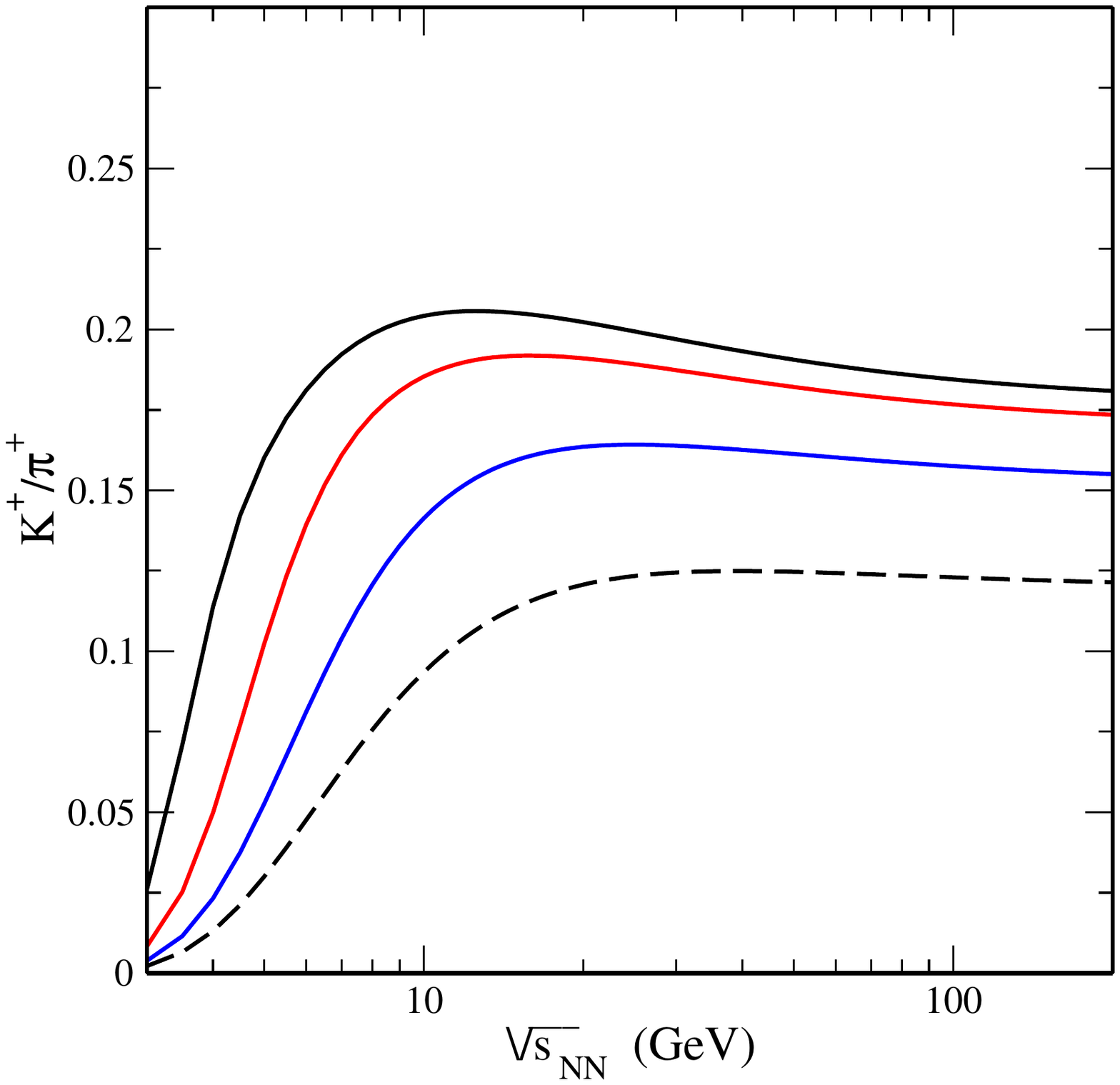} 
\includegraphics[width = 0.5\textwidth]{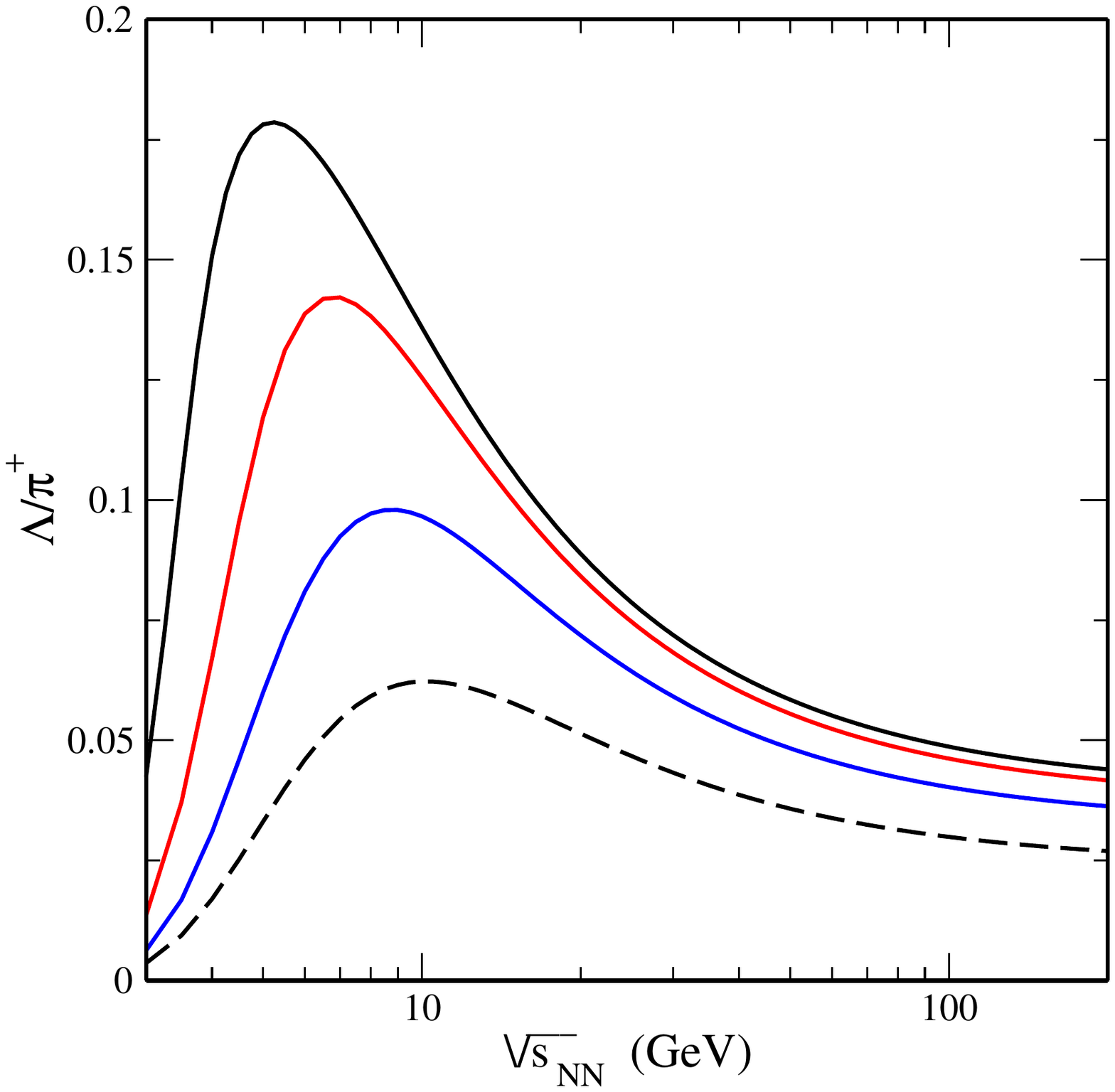} 
\caption{Values of the \kap/\pip (left-hand pane) and the $\Lambda/\pi^+$ (right-hand pane) ratios as a function
of invariant beam energy 
for various 
strangeness correlation radii $R_c$, calculated using the  
thermal model\cite{Wheaton:2004qb}. 
The correlation radius is varied from  3.0 (top curve) to 2.5, 2.0, 1.5 and finally 1.0 fm (bottom curve).
Note that the $\Lambda/\pi^+$ ratio is the ratio where the maximum stays most pronounced as the system size is 
reduced.}
\label{ratio-p}
\end{figure}

In Fig.~\ref{ratio-p}  we show the energy and system size dependence of two particle ratios calculated along
the chemical freeze-out line.
In Fig.~\ref{ratio-p}
a maximum is seen in the \kap/$\pi^+$ ratio which gradually disappears when the correlation radius
 decreases. 
A different  effect is seen in   $\Lambda/\pi^\pm$ ratio. Here, the gradual decrease
of the maximum is also seen  but, contrarily to the K$^+/\pi^+$ ratio, it  remains   quite prominent
even for a small correlation radius. 
Also, the maximum shifts, for smaller systems, towards higher \sNN. For pp collisions which 
correspond to a $R_C$ of about 1.5 fm~\cite{Kraus:2007hf}, they will hardly be observed.
It should also be noted that in the thermal model the maxima happen at  different beam energies.

\vspace{-3.5cm}
\vspace*{-0.2cm}
It must be emphasized that the results presented here are of a qualitative nature. In particular there could be changes
due to  variations with the system size of the temperature and the baryon chemical potential. In addition the strangeness equilibration volume $V_c$ could be energy dependent and
system size dependent.

\section{Conclusions}

The thermal model qualitatively describes the presence of maxima in the \kap/\pip and the $\Lambda/\pi^\pm$ ratios
at a beam energy of \sNN $\approx$ 10  GeV.  
In this talk we have
described what could possibly happen 
with different strange particles and pion yields in collisions of smaller systems due to constraints imposed by 
exact strangeness conservation. 
In particular, the $\Lambda/\pi^+$ ratio still shows a clear maximum even small systems.  The pattern of these maxima is 
also quite special as they are not always at the same beam energy.
\\
{\bf Acknowledgments}\\
K.R. acknowledges the
supported by the National  Science  {Center}, Poland  under grant Maestro,
DEC-2013/10/A/ST2/00106

\end{document}